\documentstyle[11pt,newpasp,twoside,epsf]{article}
\def\edcomment#1{\iffalse\marginpar{\raggedright\sl#1\/}\else\relax\fi}
\reversemarginpar

\begin{document}
\title{Optical Surveys for Galaxy Clusters}
\author{H.K.C.~Yee and M.D.~Gladders}
\affil{Department of Astronomy and Astrophysics, University of Toronto,
Toronto, ON M5S 3H8, Canada}
\begin{abstract}
Optical surveys for galaxy clusters have been the major
method for creating large cluster catalogs.
We give a brief review of the history of optical searches of
galaxy clusters, starting from that of Abell.
The traditional application of this survey method suffers
from contamination due to projection of galaxies along the line of
sight, which becomes increasingly more severe at higher redshift.
The new generation of wide-field CCD imagers has provided a
renewed impetus for optical surveys for clusters.
We describe a new cluster finding technique using
the red sequence of early-type galaxies in galaxy groups and clusters,
which eliminates the projection problem by essentially producing
a 3-D distribution of red galaxies using two-filter imaging data.
The Red-Sequence Cluster Survey (RCS) is a 100 square degree optical
survey, carried out using 4m class telescopes, which is optimally designed 
to search for clusters at $0.5<z<1.4$\ utilizing this technique.
We present preliminary results which indicate that the RCS is extremely
efficient in detecting galaxy clusters at these redshifts, including a number
of clusters with multiple strong lensing arcs.

\end{abstract}
\section{Introduction}
Galaxy clusters have long played an important role in both the study of
galaxy evolution and the determination of cosmological parameters.
In the former, galaxy clusters offer a unique laboratory for 
investigating the relationship between galaxy formation and evolution
and the environment; in the latter, the mass-to-light ratio and the
evolution of the mass spectrum of clusters allow us to measure
the mass density parameter, $\Omega_m$, and the amplitude of
the perturbation spectrum, $\sigma_8$ (Oukbir \& Blanchard 1992).
In both subjects, the leverage and advantage offered by having a 
sample with a large redshift coverage are clear.
It is then somewhat surprising that until recently, the largest and
most studied sample of galaxy clusters remains the Abell Catalog,
which has few clusters at $z>0.2$.
The number of known high-redshift galaxy clusters, even just at $z>0.7$,
remains small, and very often were found either as the tail of the
distribution of a systematic survey, or by serendipity.

This lack of well-defined, large, high-redshift samples is the result of
the difficulties encountered at high redshift by the two major methods
of cluster searches: the projection contamination problem in optical
surveys, and the simple distance-induced decrease of apparent flux
in X-ray searches.
However, with recent advances in technology, we are poised at an
era in which well-defined  high-redshift galaxy cluster samples, 
to a redshift well above 1, will soon become available.
This can be attributed to the recent launching of 
powerful X-ray telescopes, such as the XMM, the advent of large
format CCD imagers on 4m class optical telescopes, and the
building of a number of instruments in the millimeter wavelength regime
which will allow for blank field searches of clusters via
the Sunyaev-Zel'dovich effect.

This paper presents an abridged review of the methods and history 
for optical cluster surveys, including the latest development
with excellent results in the detection of high-redshift clusters
using optical imaging.
We first present in Section 2 a brief summary of the
three survey methods for finding clusters.
In Section 3, we concentrate on optical survey
methods and surveys from the time of the Abell cluster catalog
and onward.
In Sections 4 and 5 we describe a new technique of identifying clusters 
using two-band optical imaging data -- the red-sequence method.
We present in Section 6 preliminary results from a large survey we are carrying
out using this method. We demonstrate that wide-field optical
imaging combined with this new cluster search technique is the most
efficient method of finding $z\sim1$ galaxy clusters.

\section{Cluster Survey Methods}
\subsection{Optical}
Finding galaxy clusters using optical images, naturally, was the first method
utilized to create cluster catalogs.
Galaxies are used as markers, 
and different methods of identifying
over-densities on the projected sky have been applied to identify 
galaxy clusters.
Some of these methods are summarized in Section 3.
Finding clusters using optical images is generally considered to suffer
from one major deficiency: the projection of galaxies along the line
of sight producing contaminating signals in the search for
over-dense regions in 3-D space.
This projection effect becomes increasingly severe at higher and
higher redshift, as the column of accumulated foreground galaxies renders
the over-density produced by a galaxy cluster all but impossible
to detect, except for the richest of clusters.
The problem of projection contamination has caused optical searches
to be considered as unreliable, and their selection functions
difficult to quantify.

\subsection{X-Ray}
The detection of galaxy clusters in X-ray in 1966 (Boldt et al.)
 produced not
only a new window into the study of properties and evolution of
galaxy clusters, but also a new method for searching for
clusters.
X-ray detection provides a clean method.
This is due to,  firstly, that galaxy
clusters are the second brightest extra-galactic X-ray sources on
the sky, and can be recognized by the fact that they are resolved
in moderate resolution.
Secondly, their emissivity is proportional to $n_e^2$, which
reduces the detectability of poor clusters and groups of galaxies
relative to the rich ones, making projection contamination a
less important issue.
(However, such projection contamination, while less severe than
that found in the traditional optical searches, does exist;
e.g., in a sample of X-ray strong Abell clusters, L\'opez-Cruz \&
Yee, 2001, find a projection rate of background clusters
 as high as 15\%.)
X-ray surveys with a large sky coverage, however, are expensive 
to carry out.
Until recently, the largest and
most utilized X-ray samples of clusters were
created using clusters discovered serendipitously
in images taken for other purposes (e.g., 
Gioia et al.~1990, Romer et al.~2000).
However, with the new generation of X-ray telescopes, starting with
ROSAT, and currently with XMM, significantly larger X-ray samples
are becoming available.
A recent review of these surveys can be found in Gioia (2000).
Nevertheless, the current crop of surveys from ROSAT, still lack
the depth to discover a significant sample of $z\sim1$ clusters.

\subsection{The Sunyaev Zel'dovich Effect}
The Sunyaev Zel'dovich (SZ, Sunyaev \& Zel'dovich 1980) effect, 
produced by the scattering of
the cosmic microwave background by the hot gas in clusters, can
be used to search for galaxy clusters (see 
various papers in these proceedings).
The SZ effect has the promise of being able to detect clusters at high
redshifts, as the detectability for identical clusters
is essentially redshift independent.
However, at this point in time, most SZ work focuses on detecting
known clusters, and no new cluster has been identified using this
method.
The technological challenge of doing large blank field surveys
using the SZ effect is currently being worked out by a number
of groups, such as AMiBA, using either close-spacing array telescopes or
large bolometer arrays to achieve the necessarily field size.

\section{A Review of Optical Cluster Surveys}

\subsection{Photographic Surveys}
Systematic surveys for galaxy clusters began when the first
major photographic survey of the sky, the Palomar Optical Sky Survey
(POSS), became available.
The Abell (1958) catalog of galaxy clusters, produced by visual examination
of the POSS, is still the most studied catalog of galaxy clusters, more
than 40 years after its compilation.
Abell's method of selecting clusters, although primitive due to the
data available and the technology of the time, nevertheless, contains 
many important considerations that must go into selecting a fair
cluster sample.
These considerations also illustrate some of the difficulties that most
optical cluster surveys face.

Abell used excess galaxy counts as the primary criterion for identifying
galaxy clusters.
These counts also form the basis for the Abell Richness Class
which has been widely used as a rough indication of the richness of
galaxy clusters.
However, the statistics of galaxy counts are affected by the depth
of sampling into the galaxy luminosity function (LF) and the counting
radius around the putative cluster (see Yee \& L\'opez-Cruz 1999 for
a discussion of robust measurements of cluster richness). 
Both of these quantities require the knowledge of the redshift of
the clusters.
Abell attempted to circumvent these problems by counting galaxies 
to a depth of $m_3+2$, where $m_3$ is the third brightest galaxy
in the cluster, and within an angular radius 
inversely proportional to a redshift parameter (the ``redshift class'')
 which he assigned to all candidates based on appearance.
Clearly, this introduces a number of difficulties in that
no quantitative magnitude measurements were made and that
the estimate of redshift was necessarily  very approximate.
(We also note that  counting limits based on a quantity such as $m_3$ 
in general
produces a number of uncertainties: First, at higher redshift, 
it is subject to confusion by projection; and second,
$m_3$, being a random draw from a galaxy LF
is a function of the richness of the cluster.)
Nevertheless, Abell was able to by-and-large construct a sample of
clusters that is essentially complete and well-defined out to
$z\sim0.1$.
However, as many as 25\% of the clusters may be significantly contaminated, 
or are simply the results of projections.
Table 1 shows the rough characteristics of clusters of different
Abell richness classes, with the velocity dispersion and mass estimates
from clusters of similar richness from the CNOC1 
cluster galaxy redshift survey (Carlberg et al.~1997).

\begin{table}
\caption{Table 1: Abell Richness Class}
\begin{tabular}{r cccc}
\hline\hline
 & Abell 0 & Abell 1 & Abell 2 & Abell 3 \\
\hline
$N_{Abell}$ & 35--45   & 45--75 & 75--125 & 125--200\\
Number     & $\sim1000$  & 1224  &383 & 6 \\
$\sigma$ ({\rm km$\,$s$^{-1}$}) & 600  & 750  & 950  & 1250 \\
Mass (10$^{14}{\rm M}_\odot $) & 2 & 4 & 9 & 15 \\
\hline
\end{tabular}
\end{table}

Other cluster catalogs from photographic data
are produced primarily by simple galaxy
over-densities on the sky.
These include the Zwicky catalog (Zwicky et al.~1968),
 also using the POSS, in which
the clusters were identified using isopleths (contours of constant
galaxy density).
While projected over-density can be applied more objectively and
is more suitable for computer automation,
it produces catalogs whose completeness is strongly dependent on
redshift, as clusters of identical richness at increasingly
higher redshifts will have a lower over-density.
Later surveys using photographic plates involved digitized images
and computer algorithms.
However, they all essentially still used projected over-density as
the identification method.
These include the Edinburgh-Durham survey (Lumsden et al, 1992)
and the APM survey (Dalton et al.~1992), both based on UK Schmidt plates.
The latter used a percolation analysis to identify clusters.

The first cluster survey amied specifically
 at identifying high-redshift clusters was the  GHO survey 
(Gunn, Hoessel, \& Oke 1986), which used photographic plates to
identify possible high-redshift cluster candidates via large
red central galaxies, and then to attempt to confirm the candidates
using deep CCD images.
However, it is difficult to assess the completeness of the sample,
because of the unconventional method used to identify clusters.

\subsection{Digital Image Surveys}

Although digital imaging became available in the early 1980's, the small
field covered by the detectors then precluded their use for any substantial
cluster survey.
The largest survey available in the literature is the PDCS
(Palomar Deep Cluster Survey, Postman et al. 1996) which used drift
scan CCD images covering $\sim 6$ square degrees of sky.
The data are relatively shallow, with a limit of $V\sim24$.
The PDCS also used a more modern cluster search technique: what is now
termed the ``matched filter'' algorithm to try to reduce the effect of
projection contamination.

A matched filter algorithm essentially attempts to use some known 
properties of clusters to assist in identifying true clusters.
In the case of the PDCS, the luminosity function and the size of
the clusters were used as the matching characteristics: i.e., over-densities
with an angular size and distribution of galaxy brightness
consistent with being a cluster at certain redshift are chosen as
candidates.
The survey identified 79 cluster candidates, including a number 
of them at $z>0.7$.
However, it appears that this method also suffers from significant
projection contamination, as high as 25--30\% (Oke et al.~1998;
Holden et al.~2000).
This is primarily due to the luminosity function of cluster galaxies
not being a sharply defined function.

Dalcanton (1996) suggested a novel technique of identifying high-redshift
clusters using optical data by looking for surface brightness fluctuations
on the sky produced by high redshift clusters too far away to be well
resolved and too faint to have individual galaxies detected.
This method has the great advantage of allowing 
 the search for high-redshift clusters using a
small telescope; however, follow-up observations are
required for confirmation, and the procedure is maximally subjected
to projection effects.
A detail description of this method, and some preliminary results 
can be found in Zaritsky et al.~(2001) in this volume.

With the availability of wide-field mosaic CCD imager, currently 
there is a slew of large digital imaging surveys that can be
used for cluster surveys, though none of them are specifically
designed for that purpose.
These include, in the 4m class, the NOAO Deep survey 
covering 18 square degrees, the NOAO cosmic shear weak lensing
survey with a total planned coverage of 28 square degrees,
the CFHT cosmic shear survey with an area of 12 square degree.
The 10000 square degrees covered by the Sloan Digital Sky
Survey will provide the most complete database for 
low-redshift clusters, with a completeness limit possibly
somewhere  between $z\sim0.3$ to 0.5.
In the remainder of the paper, we will describe a new method for
identifying galaxy clusters using optical data which specifically
resolves the problem of projection. We will also present
preliminary results from a large
area survey based on this method that we are undertaking to 
produce a catalog of clusters up to $z\sim1.4$

\begin{figure}
\plotfiddle{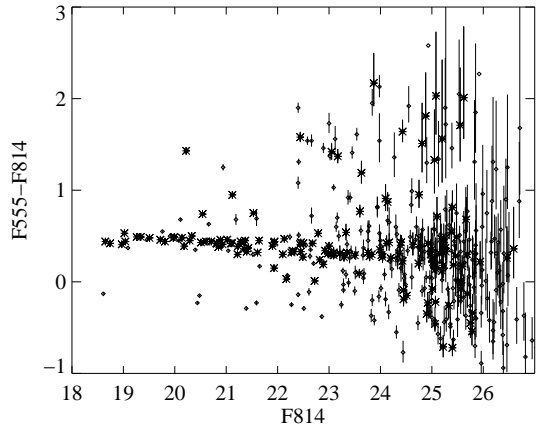}{6.cm}{0}{45}{45}{-130}{10}
\caption{An example of a red sequence of early-type galaxies in
a rich cluster (from Gladders et al.~1998).
The color-magnitude diagram of Abell 2390 ($z=0.23$)
 is based on HST images of
the cluster core with the asterisks indicating galaxies 
morphologically selected as early types.
}
\end{figure}

\section{The Red-Sequence Method}

We have developed a cluster finding method which uses a key
signature of galaxies in clusters: the color-magnitude
relation, or the red-sequence, of early-type galaxies in
clusters.
A detailed description of the cluster red sequence (CRS)
method is given in Gladders \&
Yee (2000), and we present only a summary here.

The CRS method is motivated by the observation that all rich
clusters have a population of early-type galaxies which follows
a color-magnitude relation (CMR), an example of which is shown
in Figure 1.
The red sequence represents the nominal reddest galaxy population
in any group of galaxies at the same redshift, producing a
unique signature on the color-magnitude diagram (CMD).
Furthermore, early-type galaxies dominate the bright end of
the galaxy luminosity function (LF) and the core of clusters,
or any region of high galaxy density (i.e., the morphology-density
relation, see Dressler et al.~1997).
Thus, by isolating galaxies of a specific region in the CMD
(defined by the CMR of a specific redshift), one removes most
of the galaxies accumulated in the redshift column.
Figure 2 demonstrates the red sequence for the $R-z'$ vs $z'$ CMD.

By using successive slices of CMR on the color-magnitude plane
one can effectively create a third dimension on the galaxy distribution
based on the positions of the reddest galaxies at each redshift
slice.
Overdensities (in red galaxies) can then be identified on this
three-dimensional space, selecting groups and clusters without
projection contamination.
Furthermore, the color slices can be translated into redshift slices
via either galaxy evolution models (e.g, Kodama \& Arimoto 1997)
or empirically calibrated red-sequence colors, providing an
accurate photometric redshift estimate using only two filters.

The exact implementation of the CRS method to data can vary;
Gladders \& Yee (2000) describe in detail such an implementation for 
testing the CRS method using the CNOC2 Field Galaxy Redshift Survey
database (Yee et al.~2000).
With the CNOC2 data, we are able to verify the photometrically
detected galaxy groups and clusters using the redshift data.
The test, using 1.5 square degrees of relatively shallow
data in $I$ and $V$ (5$\sigma$ detection limits of $\sim$ 22.8 and 24.0,
respectively),
showed that we are able to detect galaxy groups to a mass equivalent
to a velocity dispersion of $\sim$300 km/s out to $z=0.55$ (the
redshift completeness limit of the CNOC2 redshift catalog).
Furthermore, the photometrically estimated redshifts from the 
red sequence are accurate to an average $\Delta z$ of 0.028 (or
7\% of the mean $z$ of the sample).
We have also tested the CRS algorithm  extensively
using realistically simulated
galaxy catalogs of field and cluster galaxies.
These tests are described in Section 5 in conjunction with the
presentation of the cluster survey we carried out using the
red-sequence method.

\begin{figure}
\plotfiddle{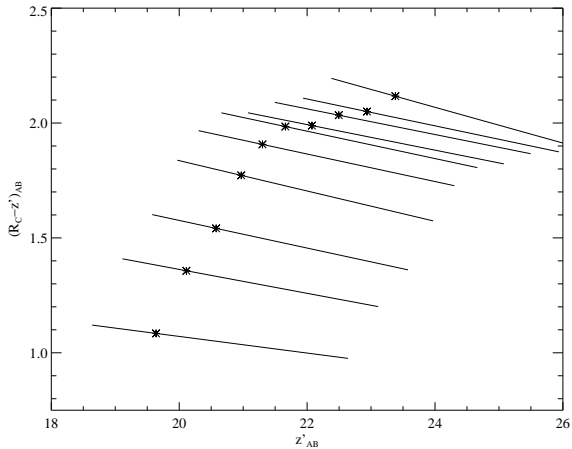}{6.cm}{0}{45}{45}{-150}{10}
\caption{Red-sequence models as a function of redshift
in $R-z'$ vs $z'$ color magnitude diagram. The red sequences
are plotted in steps of 0.1 in redshift from $z=0.5$ (bottom) to
$z=1.4$ (top).  The asterisk on each sequence
represents the $M^*$ magnitude of early-type galaxies assuming
passive evolution.
}
\end{figure}

\section{The Red-Sequence Cluster Survey (RCS)}
The efficacy of the red-sequence method, combined with the advent of
large format CCD mosaic cameras, has motivated us to carry out
an optical survey for high-redshift galaxy clusters.
In this section we briefly describe the survey and some preliminary
results.

\subsection {Survey Design}
The Red-Sequence Cluster Survey (RCS) is a 100 square degree imaging
survey using mosaic CCD cameras on 4m-class telescopes, and is
designed specifically to create a well-understood sample of
galaxy clusters to a redshift well above one.
To optimize sensitivity to high-redshift clusters we use as red a
pair of filters as possible for optical imaging: $z'$ (9200\AA)~and 
$R_C$ (6500\AA).
The $z'$ filter allows the detection of galaxies at redward of the
4000\AA~break to redshift as high  as 1.4.
The survey is divided equally between the CFHT in the northern hemisphere
using the CFH12k camera, and the CTIO 4m in the southern hemisphere
using the MOSAIC2 camera.
The former is a 12k $\times$ 8k pixel camera with a scale of 0.206$''$
per pixel, providing a field of view
of 42$'$ $\times$ 28$'$; while the latter is an 8k $\times$ 8k pixel
camera with a scale of 0.270$''$ per pixel giving a field of view 
of 36$' \times$ 36 $'$. We note that both field sizes are equivalent
to approximately 1/3 square degrees in area.

  The integration times are chosen to allow the detection of
early-type galaxies out to a redshift of 1.4 assuming a passive
evolution, and are relative short 15 and 20 minutes for $R_c$ and
$z'$ at CFHT, and 15 and 25 mins at CTIO, respectively.
These integration times provide an 8$\sigma$ ($\sim$100\%
completeness) level of $z'\sim 23.3$, and a 5$\sigma$ limit in
$R_C$ of 24.9.

The area covered by the survey is divided in a number of {\it patches},
each containing a mosaic of a number of pointings.
The pointings are slightly overlapping (by about 20$''$ to 30$''$)
to allow photometry and astrometry cross checks.
For the northern survey at CFHT, each patch contains 3$\times$5
pointings of the CFH12k camera, covering an area of approximately
2.1$^\circ \times$2.3$^\circ$.
For the southern CTIO patches, a mosaic of 3$\times$4 pointing is
used, providing a field size of approximately 1.8$^\circ\times$2.4$^\circ$.
A total of 10 patches are covered by the CFHT survey, while 12 patches
are observed for the CTIO survey.
Some of the patches are chosen to coincide with fields observed by
other surveys.
These include, for examples, the four patches of the CNOC2 Field
Galaxy Survey (Yee et al.~2000), the Groth Strip, a field from the
PDCS, and the XMM Large-Scale Structure Survey area.
The remaining patches are
 chosen based on several criteria.  They are well spaced
in Right Ascension to allow flexibility in scheduling observation time.
They have low Galactic extinction, and with galactic latitude between
45$^\circ$ and 65$^\circ$.  The upper galactic latitude bound
is applied to allow for sufficient
number of reference stars for the use of star-galaxy classification.
Fine adjustments are made in all patch placement to avoid bright
stars using the USNO-2 catalog.

  The observations are  carried out without obtaining multiply
dithered images.  There are two reasons for this strategy.
The first is simply due to the relatively short exposures,
making dithering inefficient.
Second, undithered images are inherently much easier to reduce,
as a mosaic image can be simply treated as a series of individual
 2k $\times$ 4k images, without any need to combine the different 
chips into a single image.
The small gaps between the chips will remain unfilled, but should
have no effect on the detection of clusters, which should be
many times the size of the gaps.

  Currently (as of September 2001), the northern CFHT survey is
completed, while the southern component at CTIO is about 75\% complete.
Additional $B$ and $V$ data from the CFH12k are being obtained
to augment the photometric information for the northern survey.

\subsection{Reduction and Analysis Pipelines}

Efficient pipelines are set up to automatically (with visual check
when necessary) process the images and carry out object finding,
photometry and star-galaxy classification.
The survey has an equivalent of a total of approximately 3000
2k $\times$ 4k CCD images in two colors, with about 16 million
detected objects.

Calibrated catalogs for each chip in $R_c$ and $z'$ are produced
using a modified version of the photometry package PPP (Yee 1991).
The details for the algorithms and processes for object finding,
photometry and classification can be found in Yee et al. (2000)

The individual catalogs are then astrometrically calibrated
(using M67 for internal positional calibration, and the USNO-2 catalog
for external calibration), and placed on the celestial co-ordinate
system, creating catalogs of contiguous patches.
The red-sequence cluster finding algorithm (see Gladders \& Yee 2000)
is then applied to these catalogs, locating over-density in the
three dimensional space of x-y and redshift as determined by the
red sequence color slices, creating lists of cluster candidates with estimated
red sequence photometric redshifts.

\subsection{Detectability and Completeness}
We create realistic simulated galaxy catalogs of the sky mimicking
the RCS survey to test the efficiency and selection characteristics
of the cluster finding method.
The simulations consist of realistic clusters embedded in field
galaxy catalogs which fit currently-known observed photometric
properties (magnitudes and colors) of galaxies down to $B=29$
and reproduce the known galaxy correlation function.
A wide range of cluster properties are simulated, with the primary
ones being redshift, richness, and blue galaxy fraction ($f_b$).
Figure 3 shows the detection probability of clusters of richness
of Abell classes 1 and 2 with $f_b$ of 0.45 (typical of $z>0.6$
clusters).
Similar results are obtained for simulated clusters with $f_b$ up to
0.8.
This shows that the RCS is essentially complete for  Abell class 1
out to a redshift of slightly greater than 1.
Furthermore, the simulations indicate that the false-positive rate
is low, about several per square degree, or given the expected
number density, $<5$\%, considerably smaller than conventional
optical surveys, and likely to be superior or comparable to X-ray surveys.
 
\begin{figure}
\plotfiddle{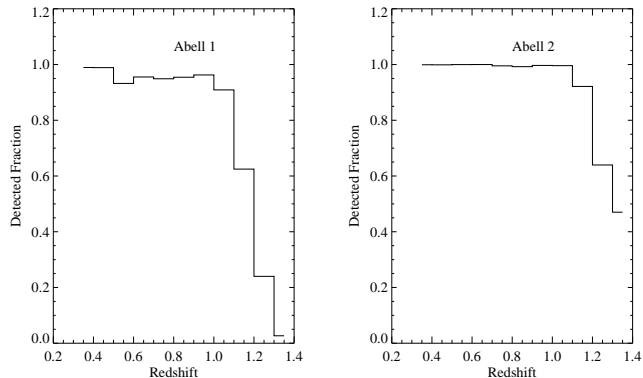}{6.cm}{0}{45}{45}{-150}{10}
\caption{
Detection probability as a function of redshift for the RCS
derived from simulated catalogs for Abell 1 and 2 richness clusters.
Both sets of simulations  have $f_b=0.45$.
}
\end{figure}

\subsection{Some Preliminary Results from the RCS}
Currently we are in the process of refining the cluster-finding
algorithm and the photometric calibration for the RCS.
Preliminary catalogs show that 
many rich clusters at $z>0.6$ have been detected.
In the following subsections, we show some of these results
which demonstrate the great richness of investigations that
can be carried out using this database.

\subsubsection{Cluster Sample:}
The RCS has been extremely efficient in detecting galaxy clusters.
In the preliminary analysis, we have discovered a significant
number of high-redshift clusters.
Figure 4 demonstrates the effectiveness of the method.
The right panel shows the smoothed surface density of total galaxy counts
in an area of $\sim$10$'\times$20$'$ in the patch RCS1620.
The middle panel illustrates the galaxy density in a $R_C-z'$ color slice
equivalent to a red sequence at $z\sim0.9$, where a significant amount
of structure is detected.
The left panel shows the gray scale $R_C$ image of the region of
the density enhancement peak, revealing a rich, compact cluster at
$z\sim0.9$.
Furthermore, the color-sliced galaxy density map shows a structure of enhanced
red galaxy density covering a linear dimension of $\sim10 h^{-1}$ Mpc,
possibly tracing out a large-scale filamentary structure.

\begin{figure}
\plotone{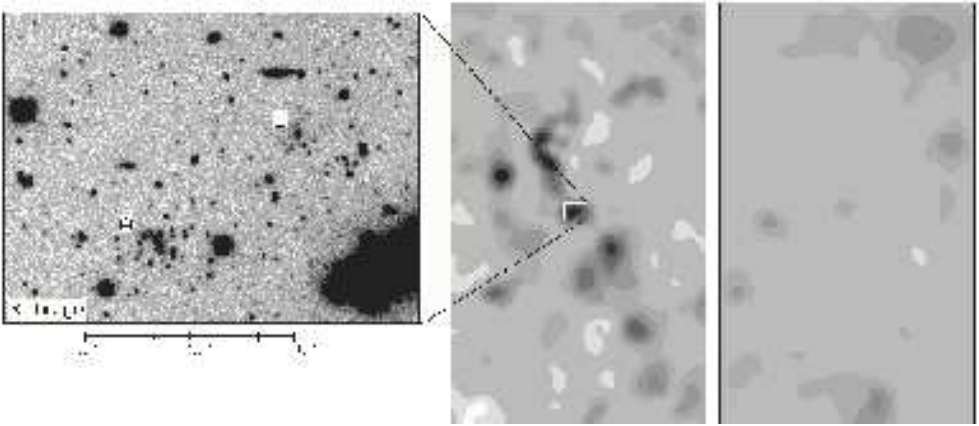}
\caption{The right panel shows the galaxy density map of all galaxies
in a 10$\times$20$'$ region in a patch from the RCS.
The central panel shows the galaxy density in a color slice 
at the color of red galaxies at $z=0.9$, revealing significant
structure of over-density.
The left panel shows a compact cluster estimated at $z=0.9$.
}
\end{figure}

Preliminary statistics from two completely calibrated and
 analysed patches from the
CFHT survey indicate that as many as 8000 galaxy groups and clusters
between redshifts of 0.2 and 1.4 will be generated by the full RCS,
with approximately 100 to 150 at $z>1$.
Figure 5 shows the number distribution of detected clusters as
a function of redshift from the two patches.
Figure 6 illustrates a small number of high-redshift rich clusters as
examples of RCS detections.

\begin{figure}
\plotfiddle{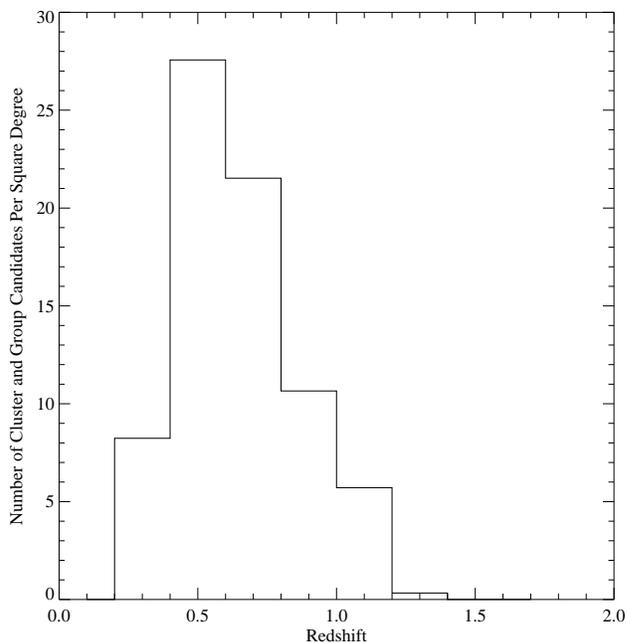}{7.8cm}{0}{45}{45}{-150}{-10}
\caption{ 
Redshift distribution (determined photometrically) of detected
clusters from two patches
of the RCS survey ($\sim 10$ square degrees).
The sample include clusters with masses down to $\sim10^{14}$ M$_\odot$.
}
\end{figure}

The photometric data alone allow us to determine a number of
physical properties of the cluster sample.
Both the luminosity function of the cluster galaxies and the
blue galaxy fraction of the clusters can be estimated using
statistical methods by correcting for background counts.
The mass can be estimated using the correlation between 
the richness parameter $B_{gc}$ (Yee \& L\'opez-Cruz, 1999)
 and velocity dispersion
found in the CNOC1 cluster sample (Yee \& Ellingson 2002).
Using properly calibrated correlations and estimated selection
functions of the survey via simulations, mass spectra of
galaxy clusters can be derived over a range between 10$^{14}$
(Abell 0) to 10$^{15}$ solar masses from redshifts between
0.3 to 1.0, providing powerful
constraints on the ($\Omega_m$,$\sigma_8$) parameter pair.
Preliminary mass functions from two patches is reported
in Gladders \& Yee (2002).

\begin{figure}
\plotone{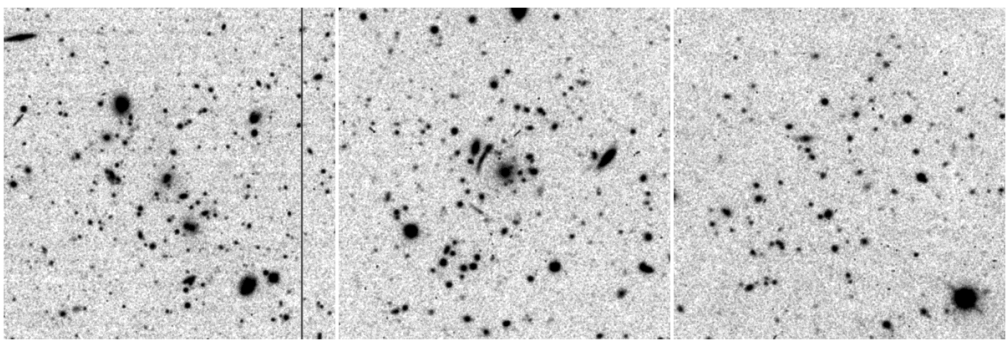}
\caption{ Gray scales of three clusters from the RCS. These are $R_C$
images with a scale of 2$'$ to a side.   Left:  A cluster at $z=0.70$;
Middle:
a cluster at $z=0.62$ with a strong arc system. 
 Right: a rich cluster
with a relatively broad spatial distribution at $z=1.0$.}
\end{figure}

\subsubsection{Gravitational Arcs:}
The RCS is designed to examine an unprecedented 
large volume of space at $z>0.5$ with sufficient depth to detect
galaxy clusters.
As a result, a number of interesting strong lensing arc systems
have been discovered.
These lensing systems in general have the lens redshift at $z\sim0.6$ to
0.8;
and, due to the relatively shallow integration, very high surface
brightness lensed arcs.
A particularly interesting system at $z=0.773$ 
(spectroscopic redshift) is shown in Figure 7.
This system has at least two large strong arcs at different redshifts
 (the first such system known).
We have obtained VLT FORS spectra of the Arc C, finding a strong
Ly$\alpha$ emission line at $z=4.879$.
Careful analysis of the structure of the emission line along
a curved slit indicates that the source has a velocity structure less
than 50 km s$^{-1}$ (Gladders, Yee, \& Ellingson, 2002).
Recent HST images show that Arc C is unresolved radially, with
preliminary models indicating a large magnification of over 30.
These data suggest that the source of Arc C is likely a single
giant star forming region or a forming spheroid.
The discovery of a large number of strong lensing clusters at
$z>0.6$ can potentially provide  powerful means of studying
high-redshift forming galaxies.

\begin{figure}
\plotfiddle{0224arc.eps}{8cm}{0}{50}{50}{-100}{0}
\caption{A rich cluster at  $z=0.773$ with bright, multiple strong arc
systems.
VLT spectroscopic data show that Arc C is at a redshift of 4.879, with
a strong Ly$\alpha$ emission line detected. }
\end{figure}

\subsubsection{Weak Gravitational Lensing:}
Although the RCS was not designed for weak lensing analysis,
the very large field coverage and the routine excellent image
quality of the data (especially those from CFHT) allow   
us to perform a number of important investigations in this subject.
Early results are summarized in Hoekstra, Yee, \& Gladders (2001a,
2001b);
these include a measurement of the cosmic shear, the bias parameter
combined or decoupled from the stochasticity parameter, and
the properties of galaxy halos.

\section{Summary}

Optical searches have been the primary and traditional method for creating
large galaxy cluster catalogs.
The problems encounted by early searches, which were based primarily on
galaxy count over-densities on the projected sky, can be overcome
by the use of color information, specifically in the form of using
the red-sequence of early-type galaxies as a marker.
This method essentially allows one to search for enhanced galaxy
density in 3-D space, removing the projection contamination problem.
The RCS has shown that such a method can produce a robust
catalog of galaxy clusters down to a mass of $10^{14} {\rm M}_\odot$
using a relatively small amount of observational
resources -- relatively shallow two-filter imaging with ground-based
4m class telescopes.
Recently, Donahue et al.~(2001) demonstrated using a small area
in the PDCS that optical search finds no evidence that optical
selection misses any X-ray luminous clusters.
Considering that
to find $7\times10^{14} M_\odot$ clusters at $z\sim 1$ requires 
the order of 10$^4$ second of integration and overhead time using
the most powerful X-ray telescope (XMM), wide-field optical cluster
search is an extremely efficient and effective method for 
creating well-defined
and well-understood high-redshift galaxy cluster catalogs.

\end{document}